\documentclass[10pt]{article}
\input epsf

\begin{document}

\title{A convolution integral representation of the thermal
Sunyaev-Zel'dovich effect}
\author{A. Sandoval-Villalbazo$^a$ and L.S. Garc\'{\i}a-Col\'{\i}n$^{b,\,c}$ \\
$^a$ Departamento de F\'{\i}sica y Matem\'{a}ticas, Universidad Iberoamericana \\
Lomas de Santa Fe 01210 M\'{e}xico D.F., M\'{e}xico \\
E-Mail: alfredo.sandoval@uia.mx \\
$^b$ Departamento de F\'{\i}sica, Universidad Aut\'{o}noma Metropolitana \\
M\'{e}xico D.F., 09340 M\'{e}xico \\
$^c$ El Colegio Nacional, Centro Hist\'{o}rico 06020 \\
M\'{e}xico D.F., M\'{e}xico \\
E-Mail: lgcs@xanum.uam.mx}
\maketitle

\begin{abstract}
Analytical expressions for the non-relativistic and relativistic
Sunyaev-Zel'dovich effect (SZE) are derived by means of suitable convolution
integrals. The establishment of these expressions is based on the fact that
the SZE disturbed spectrum, at high frequencies, possesses the form of a
Laplace transform of the single line distortion profile (structure factor).
Implications of this description of the SZE related to light scattering in
optically thin plasmas are discussed.
\end{abstract}

\section{\textbf{Introduction}}

Distortions to the cosmic microwave background radiation (CMBR) spectrum
arise from the interaction between the radiation photons and electrons
present in large structures such as the hot intracluster gas now known to
exist in the universe. Such distortions are only a very small effect
changing the brightness of the spectrum by a figure of the order of 0.1
percent. This effect, excluding the proper motion of the cluster, is now
called the thermal Sunyaev-Zel'dovich effect ~\cite{SZ1}~\cite{SZ2} and its
detection is at present a relatively feasible task due to the modern
observational techniques available. Its main interest lies on the fact that
it provides information to determine important cosmological parameters such
as Hubble's constant and the baryonic density ~\cite{rev}-\cite{Steen1}.

The kinetic equation first used to describe the SZE was one derived by
Kompaneets back in 1956 ~\cite{Peebles}~\cite{Komp}-\cite{Weymann}. This
approach implies a photon diffusion description of the effect that works
basically when the electrons present in the hot intracluster gas are
non-relativistic. Many authors, including Sunyaev and Zel'dovich themselves,
were very reluctant in accepting a diffusive mechanism as the underlying
phenomena responsible for the spectrum distortion. Later on, Rephaeli and
others~\cite{rel} computed the distorted spectrum by considering Compton
scattering off relativistic electrons. Those works show that, at high
electron temperatures, the distortion curves are significantly modified.
Useful analytic expressions that describe the relativistic SZE can be easily
found in the literature ~\cite{Sazonov-RevF}.

One possible physical picture of the effect is that of an absortion-emission
process in which a few photons happen to be captured by electrons in the
optically thin gas. An electron moving with a given thermal velocity emits
(scatters) a photon with a certain incoming frequency $\nu _{o}$ and
outgoing frequency $\nu $. The line breath of this process is readily
calculated from kinetic theory taking into account that the media in which
the process takes place has a small optical depth directly related to the
Compton parameter $y$. When the resulting expression, which will be called
\emph{structure factor}, is convoluted with the incoming flux of photons
obtained from Planck's distribution, one easily obtains the disturbed
spectra. For the non-relativistic SZE, a Gaussian structure factor has been
successfully established~\cite{yo}, but the obtention of a simple analytic
relativistic structure factor is a more complicated task~\cite{Birk}.
Nevertheless, this work shows that a relativistic structure factor can be
obtained with the help of the expressions derived in Refs.~\cite
{Sazonov-RevF} and by the use of simple mathematical properties of the
convolution integrals that describe the physical processes mentioned above.
In both cases, extensive use has been made of the intracluster gas.

To present these ideas we have divided the paper as follows. In section II,
for reasons of clarity we summarize the ideas leading to the SZ spectrum for
a non-relativistic electron gas emphasizing the concept of structure factor.
In section III we discuss the relativistic case using an appropriate
equation for the structure factor whose full derivation is still pending but
leads to results obtained by other authors using much more complicated
methods. Section IV is left for concluding remarks.

\section{General Background}

As it was clearly emphasized by the authors of this discovery in their early
publications ~\cite{SZ1}~\cite{SZ2} as well as by other authors, the
distortion in the CMBR spectrum by the interaction of the photons with the
electrons in the hot plasma filling the intergalactic space is due to the
diffusion of the photons in the plasma which, when colliding with the
isotropic distribution of a non-relativistic electron gas, generates a
random walk. The kinetic equation used to describe this process was one
first derived by Kompaneets back in 1956 ~\cite{Komp}-\cite{Weymann}. For
the particular case of interest here, when the electron temperature $Te\sim
10^{8}K$ is much larger than that of the radiation, $T\equiv
T_{Rad}\;(2.726\,\,K)$, the kinetic equation reads ~\cite{Peebles}
\begin{equation}
\frac{dN}{dy}=\nu ^{2}\frac{d^{2}N}{d\nu ^{2}}+4\nu \frac{dN}{d\nu }
\label{Uno}
\end{equation}
Here $N$ is the Bose factor, $N=(e^{x}-1)^{-1}$ , $x=\frac{h\nu }{kT}$, $\nu
$ is the frequency, $h$ is Planck's constant, $k$ Boltzmann's constant and $
y $ the ``Compton parameter'' given by,

\begin{equation}
y=\frac{k\,Te}{m_{e}c^{2}}\int \sigma _{T}\,n_{e}c\,dt=\frac{k\,Te}{
m_{e}c^{2}}\tau   \label{dos}
\end{equation}
$m_{e}$ being the electron mass, $c$ the velocity of light, $\sigma _{T}$
the Thomson's scattering cross section and $n_{e}$ the electron number
density. The integral in Eq. (\ref{dos}) is usually referred to as the
``optical depth, $\tau $'', measuring essentially how far in the plasma can
a photon travel before being captured (scattered) by an electron. Without
stressing the important consequences of Eq. (\ref{Uno}) readily available in
many books and articles ~\cite{Peebles}-\cite{imp1}, we only want to state
here, for the sake of future comparison, that since the observed value of $
N_{o}(\nu )$ is almost the same as its equilibrium value $N_{eq}(\nu )$, to
a first approximation one can easily show that
\begin{equation}
\frac{\delta N}{N}\equiv \frac{N(\nu )-N_{eq}(\nu )}{N_{eq}(\nu )}=y\left[
\frac{x^{2}e^{x}(e^{x}+1)}{(e^{x}-1)^{2}}-\frac{4xe^{x}}{e^{x}-1}\right]
\label{tres}
\end{equation}
implying that, in the Rayleigh-Jeans region $\left( x<<1\right) $,
\begin{equation}
\frac{\delta N}{N}\cong -2y  \label{cuatro}
\end{equation}
and in the Wien limit $\left( x\ >>1\right) $ :
\begin{equation}
\frac{\delta N}{N}\cong x^{2}y  \label{cinco}
\end{equation}
If we now call $I_{o}(\nu )$ the corresponding radiation flux for frequency $
\nu $, defined as
\begin{equation}
I_{o}(\nu )=\frac{c}{4\pi }U_{v}(T)  \label{seis}
\end{equation}
where $U_{v}(T)=\frac{8\pi \,h\nu ^{3}}{c^{3}}N_{eq}(\nu )$ is the energy
density for frequency $\nu $ and temperature $T$ , and noticing that
\begin{equation}
\frac{\delta T}{T}=\left( \frac{\partial (\ln I(\nu ))}{\partial
(\ln T)} \right) \left( \frac{\delta I}{I}\right)   \label{siete}
\end{equation}
we have that the change in the background brightness temperature is given,
in the two limits, by
\begin{equation}
\frac{\delta T}{T}\cong -2y\;,\;x<<1  \label{ocho}
\end{equation}
and
\begin{equation}
\frac{\delta T}{T}\cong x\,y\;,\;x>>1  \label{nueve}
\end{equation}
showing a decrease in the low frequency limit and an increase in the high
frequency one. Finally, we remind the reader that the curves extracted from
Eq. (\ref{tres}) for reasonable values of the parameter $``y"$ show a very
good agreement with the observational data.

As mentioned before, many authors, including Sunyaev and
Zel'dovich themselves, were very reluctant in accepting a
diffusive mechanism as the underlying phenomena responsible for
the spectrum distortion. The difficulties of using diffusion
mechanisms to study the migration of photons in turbid media,
specially thin media, have been thoroughly underlined in the
literature ~ \cite{z}-\cite{b}. Several alternatives were
discussed in a review article in 1980 ~\cite{imp1} and a model was
set forth by Sunyaev in the same year ~ \cite{d} based on the idea
that Compton scattering between photons and electrons induce a
change in their frequency through the Doppler effect. Why this
line of thought has not been pursued, or at least, not widely
recognized, is hard to understand. Two years ago, one of us (ASV)
~\cite{yo} reconsidered the single scattering approach to study
the SZ effect. The central idea in that paper is that in a dilute
gas, the scattering law is given by what in statistical physics is
known as the \emph{\ dynamic structure factor}, denoted by
$G(k,\nu )$ where $k=\frac{2\pi }{\lambda }$, and $\lambda $ is
the wavelength. In such a system, this turns out to be
proportional to $\exp (-\frac{(\nu - \overline{\nu} )^{2}}{W^{2}})$ where $W$ is the
broadening of the spectral line centered at frequency $\nu $
given by
\begin{equation}
W=\frac{2}{c}(\frac{2kTe}{m})^{1/2}\nu   \label{diez}
\end{equation}
If one computes the distorted spectrum through the convolution integral
\begin{equation}
I\left( \nu \right) =\int_{0}^{\infty }I_{o}\left( \bar{\nu}\right) G(k,\bar{
\nu}-(1-ay)\nu )\,d\bar{\nu}  \label{once}
\end{equation}
where the corresponding frequency shift $\frac{\delta \nu }{\nu }$ has been
introduced through Eqs. (\ref{ocho}-\ref{nueve}) $(a=-2+x)$, one gets a good
agreement with the observational data. This result is interesting from, at
least, two facts. One, that such a simple procedure is in agreement with the
diffusive picture. This poses interesting mathematical questions which will
be analyzed elsewhere, specially since the exact solution to Eq. (1) is
known (see Eq. (A-8) ref.~\cite{imp1}). The other one arises from the fact
that this is what triggered the idea of reanalyzing the SZ effect using
elementary arguments of statistical mechanics and constitutes the core of
this paper.

Indeed, we remind the reader that if an atom in an ideal gas moving say with
speed $u_{x}$ in the $x$ direction emits light of frequency $\nu _{o}$ at
some initial speed $u_{x}(0)$, the intensity of the spectral line $I\left(
\nu \right) $ is given by
\begin{equation}
\frac{I\left( \nu \right) }{I_{o}\left( \nu \right) }=\exp \left[ -\frac{
m\,c^{2}}{2kT}\left( \frac{\nu _{o}-\nu }{\nu }\right) ^{2}\right]
\label{doce}
\end{equation}

Eq. (\ref{doce}) follows directly from the fact the velocity distribution
function in an ideal gas is Maxwellian and that $u_{x}$, $u_{x}(0)$ and $\nu
$ are related through the Doppler effect. For the case of a beam of photons
of intensity $I_{o}\left( \nu \right) $ incident on a hot electron gas
regarded as an ideal gas in equilibrium at a temperature $T_{e}$ the full
distorted spectrum may be computed from the convolution integral given by
\begin{equation}
I\left( \nu \right) =\frac{1}{\sqrt{\pi }W\left( \nu \right) }
\int_{0}^{\infty }I_{o}\left( \bar{\nu}\right) \exp \left[ -\left( \frac{
\bar{\nu}-f(y)\,\nu }{W\left( \nu \right) }\right) ^{2}\right] \,d\bar{\nu}
\label{trece}
\end{equation}
which defines the joint probability of finding an electron scattering a
photon with incoming frequency $\bar{\nu}$\thinspace and outgoing frequency $
\nu $ multiplied by the total number of incoming photons with frequency $
\bar{\nu}$. $\frac{1}{\sqrt{\pi }}W\left( \nu \right) $ is the normalizing
factor of the Gaussian function for $f(y)=1$, $W\left( \nu \right) $ is the
width of the spectral line at frequency $\nu $ and its squared value follows
from Eq.(\ref{diez})
\begin{equation}
W^{2}\left( \nu \right) =\frac{4kT_{e}}{m_{e}c^{2}}\tau \,\nu
^{2}=4\,y\,v^{2}  \label{catorce}
\end{equation}
where $\tau $ is the optical depth whose presence in Eq. (\ref{catorce})
will be discussed later. The function $f(y)$ multiplying $\nu $ in Eq. (\ref
{trece}) is given by $f(y)=1+ay$ where $a=-2$ in the Rayleigh-Jeans limit
and $a=xy$ in the Wien's limit, according to Eqs. (8-9) and the fact that $
\frac{\Delta \nu }{\nu }=\frac{\Delta T}{T}$ for photons. Eq.(\ref{trece})
is the central object of this paper so it deserves a rather detailed
examination. In the first place it is worth noticing that $I_{o}(\nu )$, the
incoming flux, is defined in Eq. (\ref{seis}). Secondly, it is important to
examine the behavior of the full distorted spectrum in both the short and
high frequency limits. In the low frequency limit, the Rayleigh-Jeans limit $
\,I_{o}(\nu )$ $=\frac{2kT\nu ^{2}}{c^{2}}$, so that performing the
integration with $a=-2$ and noticing that $y$ is a very small number, one
arrives at the result
\begin{equation}
\frac{\delta I}{I}\equiv \frac{I(\nu )-I_{o}(\nu )}{I_{o}(\nu )}=-2y,\;x<<1
\label{quince}
\end{equation}
(\ref{quince}) is in complete agreement with the value obtained using Eq.(
\ref{tres}), the photon diffusion equation. In the high frequency limit
where $a=xy$ and $I_{o}(\nu )$ $=\frac{2h\nu ^{3}}{c^{2}}e^{-x}$, a slightly
more tedious sequence of integrations leads also to a result at grips with
the diffusion equation, namely
\begin{equation}
\frac{\delta I}{I}=x^{2}y,\;x>>1  \label{dieciseis}
\end{equation}
Why both asymptotic results, the ones obtained with the diffusion equation
and those obtained from Eq. (\ref{trece}) agree so well, still puzzles us.
At this moment we will simply think of them as a mathematical coincidence.
Nevertheless, it should be stressed that in his 1980 paper, Sunyaev ~\cite{d}
reached rather similar conclusions although with a much more sophisticated
method, and less numerical accuracy for the full distortion curves. The
distorted spectrum for the CMBR radiation may be easily obtained by
numerical integration of Eq. (\ref{trece}) once the optical depth is fixed, $
y$ is determined through Eq.(\ref{dos}) and $a=-2+x$. The intergalactic gas
cloud in clusters of galaxies has an optical depth $\tau \sim 10^{-2}$ ~\cite
{rel}.

From the results thus obtained in the non-relativistic case, one appreciates
the rather encouraging agreement between the observational data and the
theoretical results obtained with the three methods, the diffusion equation,
the structure factor or scattering law approach, and the Doppler effect.
This, in our opinion is rather rewarding and some efforts are in progress to
prove the mathematical equivalence of the three approaches. From the
physical point of view, and for reasons already given by many authors, we
believe that the scattering Doppler effect picture does correspond more with
reality, specially for reasons that will become clear in the last section.

\section{The relativistic case}

Following the discussion of the previous section, Eq.(\ref{trece}) may be
written as:
\begin{equation}
I\left( \nu \right) =\int_{0}^{\infty }I_{o}\left( \bar{\nu}\right) G(\bar{
\nu},\nu )\,d\bar{\nu}  \label{diecisiete}
\end{equation}
where the function $G(\bar{\nu},\nu )\,=\frac{1}{\sqrt{\pi }W\left( \nu
\right) }\exp \left[ -\left( \frac{\bar{\nu}-f(y)\,\nu }{W\left( \nu \right)
}\right) ^{2}\right] $ and $I_{o}\left( \bar{\nu}\right) $ is defined in Eq.(
\ref{seis}). It is now clear that in Wien's limit, when $\frac{h\nu }{kT}>>1$
, Eq. (\ref{diecisiete}) becomes the Laplace transform of $I_{o}\left( \bar{
\nu}\right) $ with a parameter $s=\frac{h}{kT}$ so that calling $I_{pW}$ the
distorted spectrum in that limit, one gets that:
\begin{equation}
I_{pW}(\nu )=\frac{2h}{c^{2}}\int_{0}^{\infty }\bar{\nu}^{3}\,G(\bar{\nu}
,\nu )\,e^{-s\bar{\nu}}d\bar{\nu}  \label{dieciocho}
\end{equation}
Comparing Eq. (\ref{dieciocho}) with the result obtained by integrating the
Kompaneets equation for the change of intensity in Wien's limit, namely \cite
{SZ2}-\cite{Peebles}

\begin{equation}
\Delta I=y\,\frac{2k^{3}T_{o}^{3}}{h^{2}c^{2}}x^{4}\frac{e^{x}}{\left(
e^{x}-1\right) ^{2}}\left[ -4+F\left( x\right) \right]   \label{diecinueve}
\end{equation}
where $F(x)=x\coth (\frac{x}{2})$, a straight forward inversion of the
Laplace transform yields that

\begin{equation}
G(\bar{\nu},\nu )=\delta (\bar{\nu}-\nu )-4y\frac{\nu ^{4}}{\bar{\nu}^{3}}
\delta ^{\prime }(\bar{\nu}-\nu )+y\frac{\nu ^{5}}{\bar{\nu}^{3}}\delta
^{^{\prime \prime }}(\bar{\nu}-\nu )  \label{veinte}
\end{equation}
In this equation, the primes denote derivatives with respect to $\nu $ . It
may be thus regarded as a representation of $G(\bar{\nu},\nu )$ in terms of
the Delta function an its derivatives. To illustrate this point we can show
that the RJ and Wien's limits of integral (\ref{once}) arise by substituting
$G(\bar{\nu},\nu )$ by a shifted delta function (see appendix).

In the relativistic case, the form for $G(\bar{\nu},\nu )$ may also be
written down taking the relativistic form for the particle's energy $E=\frac{
m_{o}\,c^{2}}{\sqrt{1-\frac{u^{2}}{c^{2}}}}$ and performing a power series
expansion in powers of $\frac{u^{2}}{c^{2}}$. Defining $z=\frac{k\,T_{e}}{
m_{e}\,c^{2}}$, this leads to an integral for $I(\nu )$ which reads as
follows:
\begin{equation}
I(\nu )=\frac{1}{\sqrt{\pi }W\left( \nu \right) }\int_{0}^{\infty
}I_{o}\left( \bar{\nu}\right) \exp \left[ -\frac{\sqrt{1-\frac{\left[ \bar{
\nu}-f(y)\,\nu \right] ^{2}}{2\tau y}}-1}{z}\right] \,d\bar{\nu}
\label{veintiuno}
\end{equation}
\qquad

Nevertheless, its integration even in the two limits Wien's and
Rayleigh-Jeans has been unsuccesful. However, by direct numerical
integration \ one obtains curves that agree qualitatively well with those
obtained by other methods. Thus, in order to verify the generosity of the
delta function representation we proceeded in an indirect way. It has been
shown in the literature that a good representation of the relativistic SZE
up to second order in $y$ reads as \cite{Sazonov-RevF}:

\begin{equation}
\begin{array}{l}
\Delta
I=y\,\frac{2k^{3}T_{o}^{3}}{h^{2}c^{2}}x^{4}\frac{e^{x}}{\left(
e^{x}-1\right) ^{2}}[-4+F\left( x\right) +\frac{y^{2}}{\tau
}(-10+\frac{47}{2
}F(x) \\
-\frac{47}{5}F^{2}(x)+\frac{7}{10}F^{3}(x)-\frac{21}{5}H^{2}(x)+\frac{7}{5}
F(x)H^{2}(x))]
\end{array}
\label{veintidos}
\end{equation}

Here, $H(x)=x\left[ \sinh (x/2)\right] ^{-1}$. Taking the inverse Laplace
transform in Wien's limit of Eq. (\ref{veintidos}) one obtains that
\begin{equation}
\begin{array}{l}
G_{R}(\bar{\nu},\nu )=\delta (\bar{\nu}-\nu )-\left( 4y+\frac{10y^{2}}{\tau }
\right) \frac{\nu ^{4}}{\bar{\nu}^{3}}\delta ^{\prime }(\bar{\nu}-\nu )+ \\
\left( y+\frac{47y^{2}}{2\tau }\right) \frac{\nu
^{5}}{\bar{\nu}^{3}}\delta ^{^{\prime \prime }}(\bar{\nu}-\nu
)-\frac{42\,y^{2}}{5\tau }\frac{\nu ^{6}}{ \bar{\nu}^{3}}\delta
^{^{\prime \prime \prime }}(\bar{\nu}-\nu )+\frac{
7\,y^{2}}{10\tau }\frac{\nu ^{7}}{\bar{\nu}^{3}}\delta
^{^{(IV)}}(\bar{\nu} -\nu )
\end{array}
\label{veintitres}
\end{equation}

\qquad There is of course the remaining problem, namely to obtain Eq. (\ref
{veintidos}) from the convolution integral (\ref{veintiuno}). This has so far
defied our skills, but is  purely a  mathematical problem. The main point we
want to underline here is the possibility of using expressions such as Eqs. (
\ref{veinte}) and (\ref{veintitres}), in principle obtainable from
convolution integrals with structure factors which provide a useful
analytical tool that avoids resorting too complicated and laborious methods.
In fact, Fig. 1 shows CMBR distortion comparison curves obtained using Eqs.~(
\ref{diecinueve}) and ~(\ref{diecisiete}) with (\ref{veinte}), for several
non-relativistic clusters. Fig. 2 shows the corresponding comparison
relativistic curves at higher, realistic, electron temperatures using Eqs.~(
\ref{veintidos}) and ~(\ref{diecisiete}) with (\ref{veintitres}). It is
clear that, for any practical purpose, the curves are identical. Indeed, one is not
able to identify two different plots in each figure.

\begin{figure}
\epsfxsize=3.4in \epsfysize=2.6in \epsffile{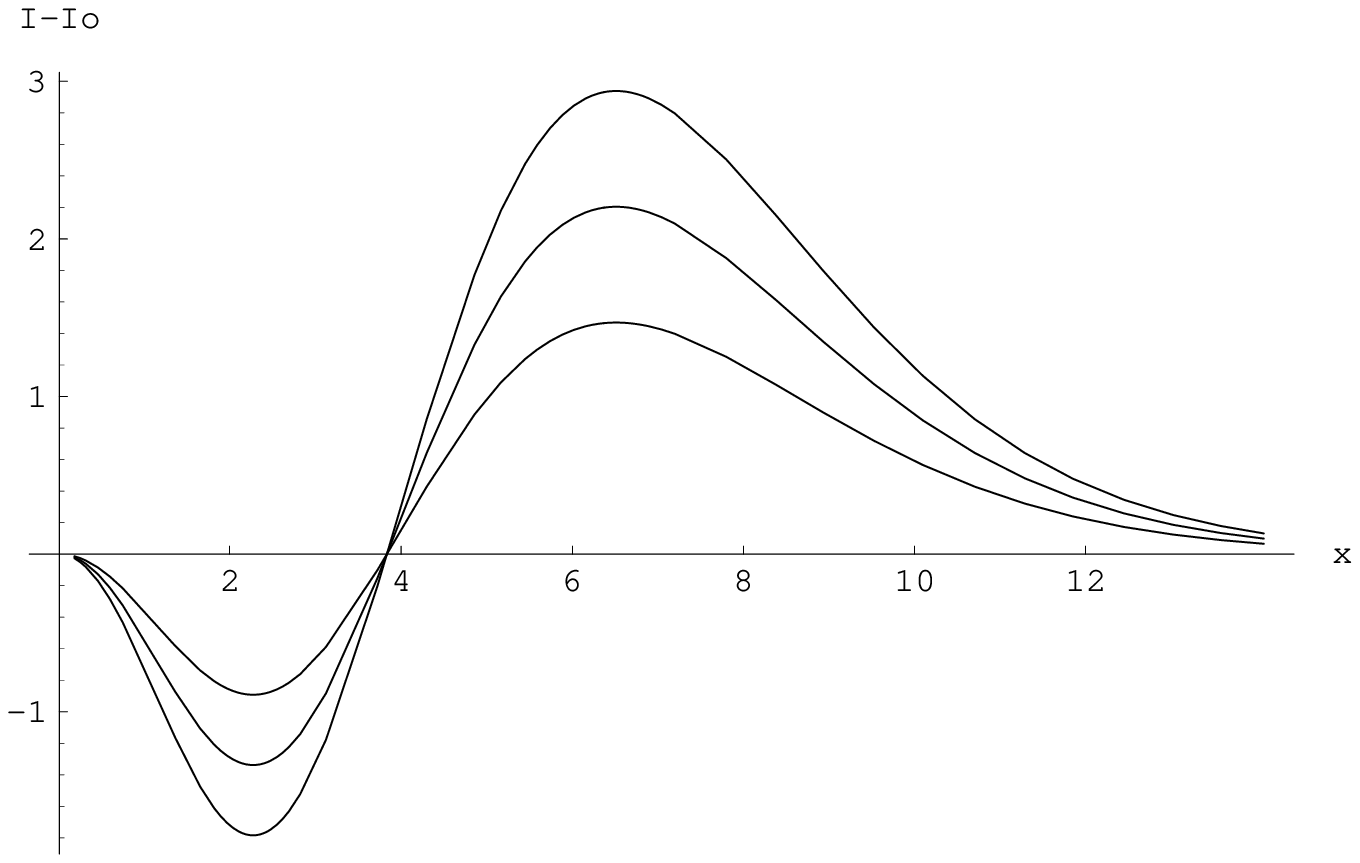}
\vspace{0.5cm} \caption{CMBR SZE distortion as
computed by Eqs. (5,10,12), with $\tau=10^{-2}$, $T=4 KeV$
(lower curve), $T=6 KeV$ (middle curve) and $T=8 KeV$ (upper
curve). No difference exists if compared the corresponding
plot of Eq. (19). $\delta I$ is measured in $erg$ $s^{-1}$
$cm^{-2}$ $ster^{-1}$. The figure is scaled by a factor of
$10^{18}$.}
\end{figure}
\vspace{0.5cm}

\begin{figure}
\epsfxsize=3.4in \epsfysize=2.6in \epsffile{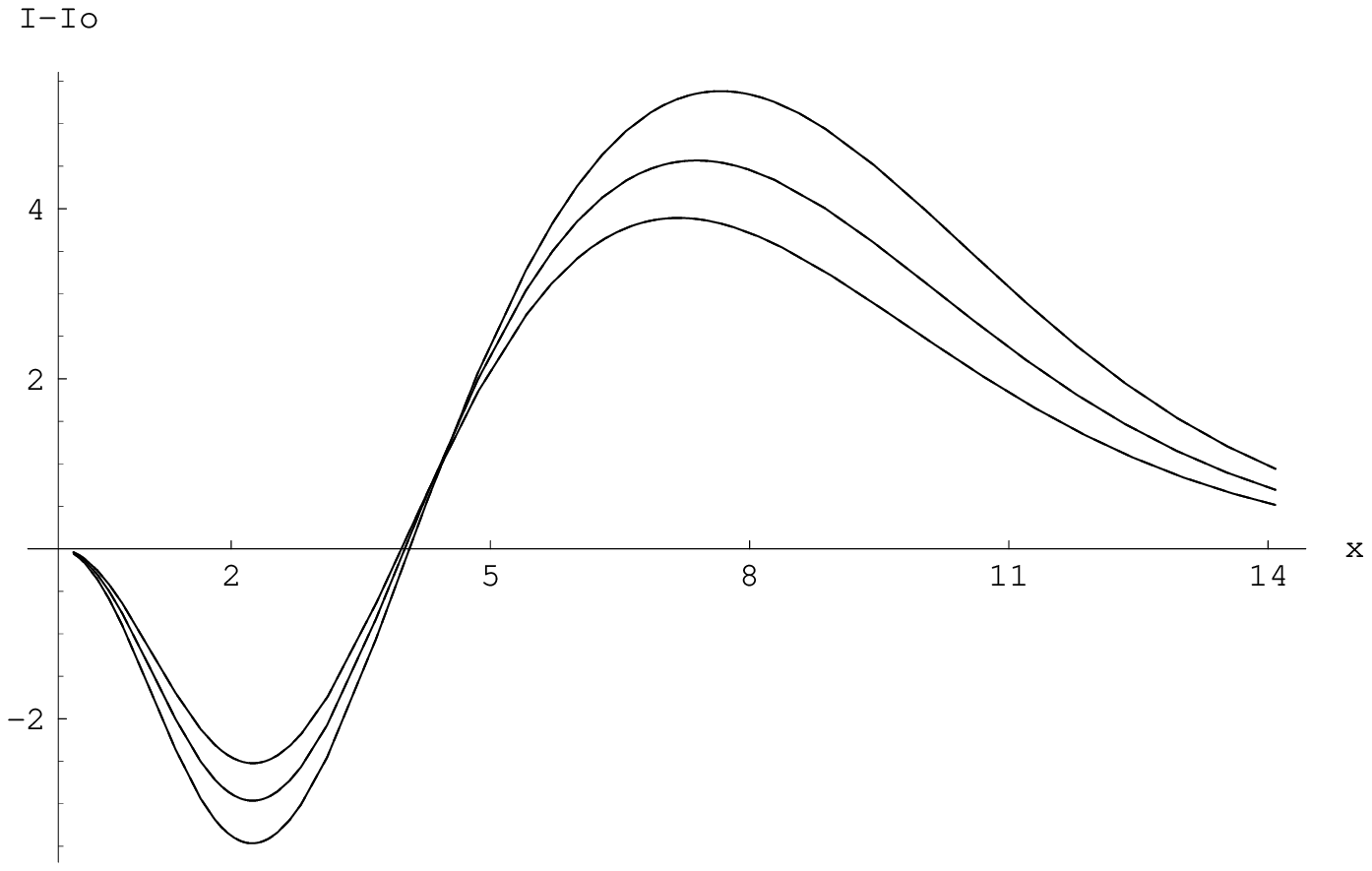}
\vspace{0.5cm} \caption{Relativistic CMBR SZE distortion as
computed by Eqs. (5,11,13), with $\tau=10^{-2}$, $T=12.5 KeV$
(lower curve), $T=15 KeV$ (middle curve) and $T=18 KeV$ (upper
curve). No difference exists if compared the corresponding
plot of Eq. (22). $\delta I$ is measured in $erg$ $s^{-1}$
$cm^{-2}$ $ster^{-1}$. The figure is again scaled by a factor of
$10^{18}$.}
\end{figure}
\vspace{0.5cm}

\section{Discussion of the results}

It is known that, for the case of a beam of photons of intensity $
I_{o}\left( \nu \right) $ incident on a hot electron gas regarded as an
ideal gas in equilibrium at a temperature $T_{e}$, the full non-relativistic
SZE distorted spectrum may be computed from the convolution integral given
in Eq. (\ref{trece}), which defines the joint probability of finding an
electron scattering a photon with incoming frequency $\bar{\nu}$\thinspace
and outgoing frequency $\nu $ multiplied by the total number of incoming
photons with frequency $\bar{\nu}$. $1-2y$ is a function that corrects the
outgoing photon frequency due to the thermal effect~\cite{imp1}. $\frac{1}{
\sqrt{\pi }}W\left( \nu \right) $ is the normalizing factor of the Gaussian,
$W\left( \nu \right) $ is the width of the spectral line at frequency $\nu $
and its squared value is defined in Eq. (\ref{catorce}).

Thus, it is clear that an accurate convolution integral between the
undistorted Planckian and a \emph{regular function} exists and can be
obtained from strictly physical arguments in the non-relativistic SZE. It is
interesting, however, that the corresponding relativistic expression does
not seem to be a simple physical generalization of Eq.~(\ref{trece}). Yet,
Eqs.~(\ref{diecisiete},\ref{veintitres}) give the correct mathematical
description of the relativistic distortion by means of Dirac delta functions
and its derivatives. Two questions can now be posed. One, of a rather
mathematical fashion, states under what conditions a structure factor such
as that appearing in Eq. (\ref{diecisiete}) can be written in a
representation involving Dirac's delta functions and its derivatives. This
subject seems to be related to the mathematical theory of distributions. The
second one, of more physical type, concerns with the possible description of
the thermal SZE as a light scattering problem in which CMBR photons interact
with an electron gas in a thermodynamical limit that allows simple
expressions for the structure factors. If this were the case, the thermal
relativistic SZE physics would not need of Montecarlo simulations or
semi-analytic methods in its convolution integral description.

Emphasis should be made on the fact that the main objective pursued in this
work is to enhance the physical aspects of the SZE by avoiding complicated
numerical techniques in many cases. This is achieved, as shown, by resorting
the concept of structure factors, or scattering laws, widely used in
statistical physics.

\bigskip

\textbf{APPENDIX}

\bigskip \qquad The starting point here is Eq.(\ref{diecisiete}). In the RJ
limit, $I_{oRJ}\,\left( \bar{\nu}\right) =\frac{2kT}{c^{2}}\bar{\nu}^{2}$.
Introducing $G(\bar{\nu},\nu )=\delta (\bar{\nu}-\nu (1-ay))$ one obtains
\[
I_{RJ}(\nu )=\frac{2kT}{c^{2}}\int_{0}^{\infty }\bar{\nu}^{2}\delta (\bar{\nu
}-\nu (1-ay))d\bar{\nu}=\frac{2kT}{c^{2}}\nu ^{2}(1-ay)^{2}
\]
keeping terms linear in $y$ we find that
\[
\frac{I_{RJ}(\nu )-I_{oRJ}\,\left( \nu \right) }{I_{oRJ}\,\left( \nu \right)
}=-2ay
\]
This result is consistent with Eq. (8) setting $a=1$.

Now, in the Wien limit, $I_{oW}\,\left( \bar{\nu}\right)
=\frac{2h}{c^{2}} \bar{\nu}^{3}e^{-\frac{h\bar{\nu}}{kT}}$, in
this case the introduction of $ \delta (\bar{\nu}-\nu (1-ay))$
leads to
\[
I_{W}(\nu )=\frac{2h}{c^{2}}\int_{0}^{\infty }\bar{\nu}^{3}e^{-\frac{h\bar{
\nu}}{kT}}\delta (\bar{\nu}-\nu (1-ay))d\bar{\nu}=\frac{2h}{c^{2}}\nu
^{3}(1-ay)^{3}e^{-\frac{h\nu }{kT}}e^{\frac{h\nu }{kT}ay}
\]%
Expanding $e^{-\frac{h}{kT}ay}$ and keeping terms up to first order in $y$
we obtain that%
\[
\frac{I_{W}(\nu )-I_{oW}\,\left( \nu \right) }{I_{oW}\,\left( \nu \right) }
=-6ay+axy
\]
Finally, setting $a=1$ and considering $x>>6$ in the Wien limit, we obtain
the desired result, Eq. (\ref{nueve}).

This work has been supported by CONACyT (Mexico), project 41081-F.

\end{document}